\documentstyle[prd,eqsecnum,preprint,aps,amsfonts,epsfig]{revtex}

\begin{document}
\renewcommand{\theequation}{\thesection.\arabic{equation}}
 \draft
  \title{Lower Neutrino Mass Bound from SN1987A Data and Quantum Geometry}
\author{G. Lambiase$^{a,b}$, G. Papini$^{c,d,e}$, R. Punzi$^{a,b}$, G. Scarpetta$^{a,b,e}$}
\address{$^a$Dipartimento di Fisica "E.R. Caianiello"
 Universit\'a di Salerno, 84081 Baronissi (Sa), Italy.}
  \address{$^b$INFN - Gruppo Collegato di Salerno, Italy.}
  \address{$^c$Department of Physics, University of Regina, Regina, SK, S4S 0A2, Canada.}
  \address{$^d$Prairie Particle Physics Institute, Regina, SK, S4S
  0A2, Canada}
  \address{$^e$International Institute for Advanced Scientific Studies, 89019 Vietri sul Mare (SA), Italy.}
\date{\today}
\maketitle
\begin{abstract}
A lower bound on the light neutrino mass $m_\nu$ is derived in the
framework of a geometrical interpretation of quantum mechanics.
Using this model and the time of flight delay data for neutrinos
coming from SN1987A, we find that the neutrino masses are bounded
from below by $m_\nu\gtrsim 10^{-4}-10^{-3}$eV, in agreement with
the upper bound $m_\nu\lesssim$ $\left({\cal O}(0.1) - {\cal O}
(1)\right)$ eV currently available. When the model is applied to
photons with effective mass, we obtain a lower limit on the
electron density in intergalactic space that is compatible with
recent baryon density measurements.
\end{abstract}

\vspace{0.3in}


Keyword(s):Quantum Geometry, Other topics on General Relativity,
Neutrino mass, Supernovae

\section{Introduction}
\setcounter{equation}{0}

The detection of neutrinos from SN1987A by the Kamiokande and IMB
experiments has confirmed the stellar collapse model, giving new
insights into neutrino physics.

In essence, the experimental results are as follows. The first
optical record of the supernova SN1987A occurred at $\sim
10:37:55$ UT on 23 February 1987 \cite{shelton}. About three hours
before, neutrino signals were observed, almost simultaneously, at
the Kamioka (7:35:35 UT) and IMB (7:35:41 UT) detectors
\cite{hirata}.

The small difference between the arrival times of neutrinos and
photons (on a distance of $52\pm5$ Kpc between SN1987A and Earth)
was used to set some constraints on the validity of the
equivalence principle \cite{longo}, by attributing the time delays
undergone by photons and neutrinos to different gravity couplings.
This analysis rests, however, on the hypothesis that the time
difference of the emission of photons and neutrinos during the
explosion is not very large.

In the case of massive particles, the signals from SN1987A were
also used to evaluate an upper bound on their masses
\cite{arnett}. If, in fact, $d$ is the distance of the neutrino
source from Earth, then the time of flight delay of neutrinos
relative to that of photons emitted by the same source is
\begin{equation}\label{flight-time}
  \Delta t=\delta t_\nu-\delta t_\gamma\simeq \frac{m_\nu^2}{2E^2}\, d\,.
\end{equation}
There is no unanimous agreement on how to fix an upper limit on
the duration of neutrino emission. Old analyses of the Kamiokande
data use $\sim 4 s$ as such a limit and yield an upper bound
$m_{\nu}\lesssim 27 eV$ or $m_{\nu}\lesssim 12eV$, depending on
the choice of events used. On the contrary, other models of
supernova explosion allow for different limits, of order $\sim 10
s$ \cite{raffelt}. Consequently, we will consider a varying
emission time limit in the following analysis and derive our
results as a function of this parameter.

The aim of this work is to derive a lower bound on the neutrino
mass in the framework of a model introduced by Caianiello and
co-workers to provide quantum mechanics with a geometrical
framework \cite{cai3}. An interesting consequence of the model is
that the proper acceleration of massive particles has an upper
limit ${\cal A}_m = 2mc^{3}/\hbar$, where $m$ is the invariant
rest mass of the particle. This mass dependent limit, or maximal
proper acceleration (MA) ${\cal A}_{m}$, can be derived from
quantum mechanical considerations \cite{ca,pw,papinix} and the
fact that the acceleration is largest in the particle rest frame.
The absolute value of the proper acceleration therefore satisfies
the inequality $ a \leq {\cal A}_m $. No counterexamples are known
to the validity of this inequality that has at times been elevated
to the status of principle.

Classical and quantum arguments supporting the existence of a MA
have been discussed in the literature \cite{prove,wh,b}. MA also
appears in the context of Weyl space \cite{pap}, and of a
geometrical analogue of Vigier's stochastic theory \cite{jv} and
plays a role in several issues. It is invoked as a tool to rid
black hole entropy of ultraviolet divergences \cite{McG}. MA is at
times regarded as a regularization procedure \cite{nesterenko}
that avoids the introduction of a fundamental length \cite{gs},
thus preserving the continuity of space-time.

An upper limit on the acceleration also exists in string theory
where Jeans-like instabilities occur \cite{gsv,gasp} when the
acceleration induced by the background gravitational field reaches
the critical value $a_c = \lambda^{-1} = (m\alpha)^{-1}$ where
$\lambda$, $m$ and $\alpha^{-1}$ are string size, mass and
tension. At accelerations larger than $a_c$ the string extremities
become casually disconnected. Frolov and Sanchez \cite{fs} have
found that a universal critical acceleration must be a general
property of strings. It is the same cut--off required by Sanchez
in order to regularize the entropy and the free energy of quantum
strings \cite{sa2}.

Applications of Caianiello's model include cosmology \cite{infl},
the dynamics of accelerated strings \cite{Feo} and neutrino
oscillations \cite{8,qua}. The model also makes the metric
observer--dependent, as conjectured by Gibbons and Hawking
\cite{Haw}.

Recently, the model has been applied to particles falling in the
gravitational field of a spherically symmetric collapsing object
\cite{sch}. In this problem MA manifests itself through a
spherical shell external to the Schwarzschild horizon and
impenetrable to classical and quantum particles \cite{boson}. The
shell is not a sheer product of the coordinate system, but
survives, for instance, in isotropic coordinates. It is also
present in the Reissner-Nordstr\"om \cite{reiss} and Kerr
\cite{kerr} cases. In the model, the end product of stellar
collapse is therefore represented by compact, impenetrable
astrophysical objects whose radiation characteristics are similar
to those of known bursters \cite{papiniz}.

Caianiello's model is based on an embedding procedure \cite{sch}
that stipulates that the line element experienced by an
accelerating particle is represented by
\begin{equation} \label{eq1}
d\tau^2=\left(1+\frac{g_{\mu\nu}\ddot{x}^{\mu}\ddot{x}^{\nu}}{{\cal
A}_m^2}
\right)g_{\alpha\beta}dx^{\alpha}dx^{\beta}=\left(1+\frac{a^2(x)}{{\cal
A}_m^2}
 \right) ds^2\equiv
\sigma^2(x) ds^2\,,
\end{equation}
where $g_{\alpha\beta}$ is a background gravitational field. The
effective space-time geometry given by (\ref{eq1}) therefore
exhibits mass-dependent corrections that in general induce
curvature and violations of the equivalence principle. The MA
corrections appear in the conformal factor in (\ref{eq1}) and can
not therefore modify null geodesics, hence the dynamics of
massless particles, as stated below.

The four--acceleration $\ddot x^\mu = d^2 x^\mu/d\,s^2$ appearing
in (\ref{eq1}) is a rigorously covariant quantity only for linear
coordinate transformations. Its transformation properties are
however known and allow the exchange of information among
observers. Lack of covariance for $\ddot x^\mu$ in $\sigma^2(x)$
is not therefore fatal in the model. The justification for this
choice lies primarily with the quantum mechanical derivation of MA
which applies to $\ddot x^\mu $, requires the notion of force, is
therefore Newtonian in spirit and is fully compatible with special
relativity. The choice of $\ddot x^\mu $ in (\ref{eq1}) is, of
course, supported by the weak field approximation to $g_{\mu\nu}$
which is, to first order, Minkowskian. On the other hand,
Einstein's equivalence principle does not carry through to the
quantum level readily \cite{lamm,singh}, and the same may be
expected of its consequences, like the principle of general
covariance \cite{wein}. Complete covariance is, of course,
restored in the limit $\hbar \to 0$, whereby all quantum
corrections, including those due to MA, vanish.

As shown below, the existence of a lower bound on the neutrino
mass follows, in this framework, from the fact that MA corrections
depend inversely on the mass of the particle (as well as directly
on their energy). In particular, the gravitational time delay
undergone by neutrinos in the interaction with the gravitational
field of the supernova is affected by these corrections.

In Section II, we consider the data registered by the Kamiokande
experiment, and perform an analysis similar to that used in
\cite{arnett} and briefly mentioned above. Moreover, we assume, as
in \cite{arnett}, that all events are due to neutrinos of the same
mass and we do not consider the possibility of neutrino
oscillations.

Even though our method can also be applied to the photon-neutrino
delay, the comparison between the arrival times of neutrinos of
different energies provides a more stringent bound.

The time delays are calculated in the Section III, which is then
followed by the actual determination of the lower mass bounds and
a short discussion.

\section{Time delays and maximal acceleration}
\setcounter{equation}{0}

It is sufficient, for our purposes, to consider particles that
escape {\it radially} from the SN core. The MA effects induced by
the gravitational field of our Galaxy are completely negligible,
even though the time delay caused by our Galaxy is relevant in
testing the equivalence principle \cite{longo}. In the following
we will use units  $ \hbar =c =1$.

The time delay for a massive particle is given by \cite{wh}
\begin{equation}\label{shapiro1}
  \delta t = \left(1+\frac{m^2}{2E^2}\right) (d-r)+
  \frac{1}{2}\int_{r}^{d} h_{\mu\nu}(r')k^\mu k^\nu dr'\,,
\end{equation}
where the metric deviation $h_{\mu\nu}=g_{\mu\nu}-\eta_{\mu\nu}$
for a weak, spherically symmetric gravitational field
characterized by the Newtonian potential $\phi(r)$, is given, in
General Relativity, by $h_{00}=2\phi(r)$, $h_{ij}=-2\phi(r)
\delta_{ij}$. In (\ref{shapiro1}) the particle momentum is
$k^\mu=(1, {\hat {\bf k}})$ $({\hat {\bf k}}={\bf k}/E)$, so that
$k^\mu k_\mu = m^2/E^2 $. The point where the particles are
generated is  $r\ll d $. The reference frame is located at the
source center.

In order to compute the MA corrections to the time delay we start
with the effective metric (\ref{eq1}) experienced by a massive
particle and write
\begin{equation}\label{metric}
  g_{\mu\nu}=\eta_{\mu\nu}+h_{\mu\nu}+\frac{a^2}{{\cal A}_m^2}\,
  \eta_{\mu\nu}+{\cal O}\left(\frac{a^2}{{\cal A}_m^2}\,
  \,h_{\mu\nu}\right)=\eta_{\mu\nu}+h_{\mu\nu}+{\tilde h}_{\mu\nu}\,.
\end{equation}
In the ultra-relativistic approximation $E\gg m$
\cite{wh,sch,boson} we find
\begin{equation}\label{a^2}
  {\tilde
  h}_{\mu\nu}(r')k^\mu k^\nu=\frac{m_\nu^2}{{\cal A}_m^2E^2}\eta_{\mu\nu}{\ddot x}^\mu{\ddot x}^\nu\simeq
  \frac{4}{{\cal A}_{m}^2}\left(\frac{E}{m_\nu}\right)^2\phi'^2(r)=
  \frac{4}{{\cal A}_{m}^2}\left(\frac{E}{m_\nu}\right)^2\left(\frac{GM_{r}}{r^2}\right)^2,
\end{equation}
where $M_{r}$ is the mass enclosed in the sphere of radius $r$
concentric to the gravitational source.

The corresponding result for rigorously massless photons cannot be
obtained from (\ref{a^2}) in the limit $m\to 0$. It is shown in
the Appendix that for photons ${\tilde h}_{\mu\nu} k^\mu k^\nu=0$,
and that, therefore, their MA corrections do not contribute to the
present problem.

With the exclusion of massless particles, the MA corrections
(\ref{a^2}) scale as $1/m^4$. In fact, in the weak field and
ultrarelativistic approximations, the proper acceleration $a$ is
related to the acceleration in the lab frame $\ddot{r}$ by
$|a|^2=\gamma^2|\ddot{r}|^2\simeq\gamma^4\phi'^{2}(r)$, where
$\gamma=E/m$. The $m^{-4}_{\nu}$ dependence is then produced by
the factor $(m/E)^2$ due to the definition of $k^{\mu}$ and the
factor $1/m^2$ due to ${\cal A}^{-2}_{m}$.

Eq.(\ref{shapiro1}) becomes
\begin{equation}\label{shapiroMA}
  \delta t_{{\cal A}_m}=\left(1+\frac{m_\nu^2}{2E^2}\right)(d-r)+\frac{1}{2}\int_{r}^{d}{\tilde
  h}_{\mu\nu}(r')k^\mu k^\nu dr' =
  \delta t +\frac{2}{{\cal A}^2_m}\left(\frac{E}{m_{\nu}}\right)^2\int_{r}^{d}
  \phi'\,^2(r') dr'\,.
\end{equation}

In order to perform the integral in (\ref{shapiro1}) and obtain
the MA correction to be compared with the experimental data, we
need a model of the supernova and of its gravitational field. For
the purpose of estimating a lower bound for the neutrino mass,
very detailed models are not particularly useful. To give
significance to our results, it is nevertheless necessary to show
that they are not too dependent on the model used to describe the
collapsing star. We therefore use in the following a simple model
and show that the results are sufficiently robust against
reasonable variations of the parameters. The model we refer to is
intended to describe the density profile of the supernova a few
milliseconds after the implosion, when, according to our current
theoretical understanding, neutrinos are emitted \cite{hille}. We
consider a time-independent mass distribution, implicitly assuming
that the time scale of the emission is so rapid that the evolution
of the profile is negligible. We choose a density profile
\begin{equation}
  \rho(r)=\left\{ \begin{array}{ll} \rho_c &  \quad r<r_c  \\
          \rho_c\left(\displaystyle{\frac{r}{r_c}}\right)^{-n} &
          \quad
          r>r_c\,,  \end{array} \right. \label{profile}
\end{equation}
which describes a SN with a hard core and a halo of matter of
decreasing density. As typical parameters we take $\rho_c=10^{14}$
g cm$^{-3}$, $r_c=15$ Km, $3.5 \leq n \leq 10$. Their influence on
the final results is discussed later. An elementary integration
gives
\begin{equation}\label{m(r)}
M_{r}= \left\{ \begin{array}{ll} M_c\left(\displaystyle{\frac{r}{r_c}}\right)^3 & \quad r<r_c \vspace{0.2in} \\
     M_c\left[\displaystyle{\frac{n-2}{n-3}-\frac{1}{n-3}\left(\frac{r_c}{r}\right)^{n-3}}\right] & \quad r>r_c
     \,,
     \end{array} \right.
\end{equation}
where $M_c\simeq 1.4\,\,10^{30}\mbox{Kg}\simeq 0.7\, M_\odot$ is
the core mass. Our model satisfies the weak field approximation to
the gravitational field and the perturbational approach required
to calculate the MA corrections. As reference values, we have, for
$r=r_c$,
\begin{eqnarray}\label{weakfield}
  \phi(r_c)&=&\frac{n-1}{n-2}\frac{GM_c}{r_c}\sim 0.15 < 1\\
  g_c&=&\frac{GM_c}{r_c^2}\simeq 4.15\,\,10^{11}\frac{\mbox{m}}{\mbox{s}^2}\ll {\cal
  A}_m\quad for\quad m_\nu\gg 10^{-12}\mbox{eV}\nonumber\,.
\end{eqnarray}
Both approximations therefore hold true.

As our current knowledge of SN dynamics suggests, we assume that
neutrinos are generated outside the core, a distance $R$ from the
centre. Because $d\gg R$, we can also replace the upper limit of
integration in (\ref{shapiroMA}) with $+\infty$. We then have
\begin{eqnarray}\label{aint}
  \int_R^\infty \phi'\,^2(r') dr' &=& \frac{g_c^2r_c}{(n-3)^2}\int_x^\infty \frac{\left(n-2-z^{3-n}\right)^2}{z^4}dz \\
  &=&
  \frac{g_c^2r_c}{(n-3)^2}\left[\frac{(n-2)^2}{3x^3}-\frac{2(n-2)}{nx^n}+\frac{1}{(2n-3)x^{2n-3}}\right]\nonumber\\
  &\dot{=}& g_c^2r_c\, F(x,n)
  \nonumber\,,
\end{eqnarray}
where $x=R/r_c$. Considering the substantial agreement among the
current SN models on the values of $M_c$ and $r_c$, the function
$F(x,n)$ is the really model-dependent part of our calculation. It
could in principle be very sensitive to the choice of the
parameters and must therefore be analyzed carefully. According to
current astrophysical models, the choice $1.5\leq x \leq 5.5$,
i.e. 20 Km $\leq R\leq $ 80 Km, is plausible. To have an idea of
the magnitude of $F(x,n)$ and of its variation, we plot it in
Fig.\ref{F(x,n)} for values of the parameters in the range
considered. We see that $0.0025\lesssim F(x,n)\lesssim 0.25$. For
the sake of clarity, we write $F(x,n)=0.025\,\Delta$, with
$1/10\leq\Delta \leq10$. The uncertainty in $F(x,n)$ extends then
by two orders of magnitude. We will see, however, that this
results only in a very small error in the estimate of the neutrino
mass.

Finally, the time delay for a massive particle with MA corrections
can be written in the form
\begin{equation}\label{timedel}
  \delta t_{{\cal A}_m}=\delta t + 2r_c\left(\frac{g_c}{{\cal
  A}_m}\right)^2\left(\frac{E}{m_\nu}\right)^2F(x,n)\,,
\end{equation}
or, numerically, as
\begin{equation}\label{timedel2}
  \delta t_{{\cal A}_m}\simeq\delta t
  + 0.8\,\, 10^{-16}\Delta\left(\frac{E}{\mbox{MeV}}\right)^2\left(\frac{m_\nu
  }{\mbox{eV}}\right)^{-4}\;\mbox{s}\,.
\end{equation}

If we re-write the last expression as
\begin{equation}\label{timedele}
  \delta t_{{\cal A}_m}\simeq\delta t
  + 0.8\,\, 10^{-16}\left(\frac{E}{\mbox{MeV}}\right)^2\left(\frac{1}{\sqrt[4]{\Delta}}\frac{m_\nu
  }{\mbox{eV}}\right)^{-4}\;\mbox{s}\,,
\end{equation}
we see that an error of one order of magnitude in $F(x,n)$ only
results in a factor $\sqrt[4]{10}\simeq 1.8$ in our estimate of
the neutrino mass. This clearly shows the stability of our
predictions against a reasonable variation of the parameters used.

Starting from this result, we derive in the next section a lower
bound on the neutrino mass, using the experimental results on the
relative time delay between neutrinos of different energies.

\vspace{0.15in}
\section{Lower bounds on the neutrino mass}
\setcounter{equation}{0}

The arrival time of neutrinos of different energies, recorded by
the Kamioka experiment, can be now analyzed. We closely follow the
standard analysis that leads to an upper bound on the mass of the
neutrinos \cite{arnett}.

If $t_0\simeq5.3\,\,10^{12}$ s is the light travel time from
SN1987A to Earth for a neutrino of energy $E$, then the
relationship between observation and emission times is
\begin{equation}\label{obs-em}
  t_{obs}-t_{em}=t_0\left(1+\frac{m_{\nu}^2}{2E^2}\right)+\delta t_{{\cal
  A}_m}\,.
\end{equation}
As above, we assume that $t_{0}m_{\nu}^2/2E^2$ is negligible with
respect to $\delta t_{{\cal A}_m}$ in the mass range of interest,
and discuss this assumption at the end of our analysis. We then
obtain
\begin{equation}\label{em-obs}
  t_{em}=t_{obs}-\left[0.8\,\, 10^{-4}\left(\frac{E}{\mbox{MeV}}\right)^2\left(\frac{m_\nu}
  {\sqrt[4]{\Delta}10^{-3}\mbox{eV}}\right)^{-4}\right]s \,.
\end{equation}
As normally done in the literature \cite{arnett}, we consider only
events 1-5 and 7-9 of the Kamioka experiment (see Table I). We
exclude event 6 because it received fewer than 20 photomultiplier
hits, and events 10-12, probably associated with a late burst of
the SN or with a long tail of the emission time distribution.

\begin{figure}
\centering \leavevmode \epsfxsize=9cm \epsfysize=7cm
\epsffile{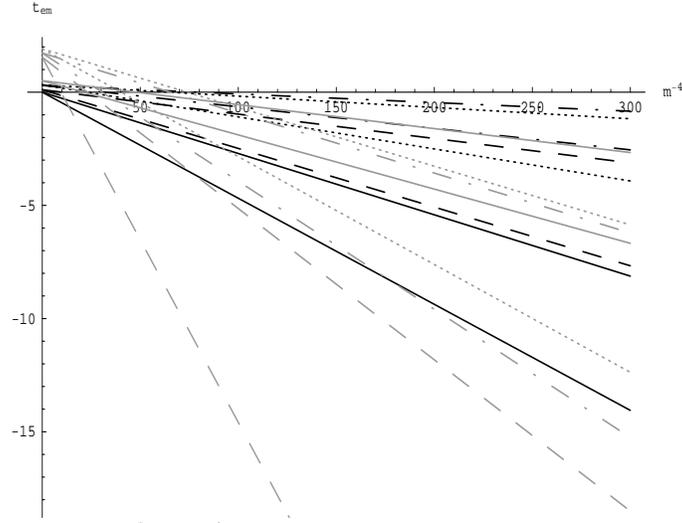} \caption{\small{$t_{em}(s)$ vs.
$(m_{\nu}/10^{-3} \mbox{eV})^{-4}$ for the eight events analyzed:
1.solid black line 2.dashed black line 3.dot-dashed black line
4.dotted black line 5.solid grey line 7.dashed grey line 8.dot
dashed grey line 9.dotted grey line.}}\label{Fig}
\end{figure}

The values of $t_{em}$ obtained from the data are shown in
Fig.\ref{Fig} as a function of $(m_{\nu}/\sqrt[4]{\Delta})^{-4}$.
For each event two lines are shown, corresponding to energies at
the upper and lower limits of the declared error in the energy
measurements.

\begin{figure}
\centering \leavevmode \epsfxsize=6cm \epsfysize=6cm
\epsffile{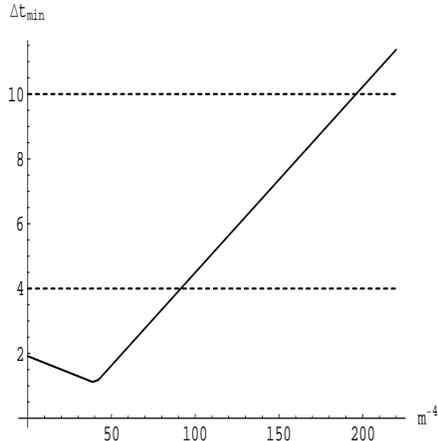} \caption{\small{Minimum time interval over
which neutrinos of the Kamioka experiment could have been emitted
as a function of $(m_{\nu}/\sqrt[4]{\Delta})^{-4}$ in unit of
$10^{-3}$ eV. }}\label{deltat}
\end{figure}

We now calculate the minimum time interval over which neutrinos
could have been emitted. The result is shown in Fig.\ref{deltat},
where this quantity is plotted as a function of
$(m_{\nu}/\sqrt[4]{\Delta})^{-4}$. By requiring that the minimum
time interval is less than a fixed value (depending on the
specific supernova model), we obtain a significant lower bound on
the neutrino mass.

As two relevant examples, we choose $\Delta t\leq 4s$ and $\Delta
t \leq 10 s$. We obtain
\begin{eqnarray}\label{delta t}
\Delta t\leq 4s
\quad\quad\quad&\quad&\left(\frac{m_{\nu}}{\sqrt[4]{\Delta}
10^{-3} {\mbox {eV}} }\right)\lesssim 91.4\quad\to\quad
m_{\nu}\gtrsim(0.17 -
0.54)10^{-3}{\mbox eV}\\
\Delta t\leq 10s
\quad\quad\quad&\quad&\left(\frac{m_{\nu}}{\sqrt[4]{\Delta}
10^{-3} {\mbox {eV}} }\right)\lesssim 196.1\quad\to\quad
m_{\nu}\gtrsim(0.15 - 0.48)10^{-3}{\mbox eV}\,.\nonumber
\end{eqnarray}

We note that our conclusions are very robust against a variation
of the supernova explosion model.

We can verify that the kinematical time delay is not relevant for
a mass of this order of magnitude. Recalling that the light travel
time is $t_0\simeq 5.3\,10^{12}\,\,\mbox{s}$, the kinematical
contribution to $\delta t$ is
\begin{equation}\label{deltatkin}
  \delta t_{k}=t_0\frac{m_{\nu}^2}{2E^2}\approx10^{-8}\mbox{s}
\end{equation}
for $E\approx 10 MeV$, which is completely negligible.

In other approaches, an improvement in the bound is obtained by
considering only events 1-5 as neutrinos emitted in an initial
pulse of less than $ 1s$ duration. This type of analysis does not
lead to a sizable improvement in our case and we do not pursue it
here.

It has been pointed out that photons acquire an effective mass
$m_{eff}$ while travelling in the intergalactic medium and that
(\ref{shapiroMA}) may therefore be applied to this physical
situation with the substitution $m_\nu\to m_{eff}$. This follows
from the fact that the dispersion relation of the photon
propagating in a medium becomes $\omega^2-k^2=\Pi_a(\omega, k)$,
where $\Pi_a$ are functions that represent the medium response to
the electromagnetic field. The effective mass is then defined by
$m_{eff}^2=\Pi_a(\omega, k)$ and differs for different
polarizations and wave vectors \cite{raffelt}. It is difficult, in
the general case, to think of $m_{eff}$ as an object capable of a
single, unified response to mechanical and gravitational
solicitations. If, however, $m_{eff}$ becomes a constant
independent of wave number and frequency as in the case of an
interstellar plasma at $T=0$, then $m_{eff}
=\omega_p=\sqrt{4\pi\alpha\, \frac{n_e}{m_e}}$, where $\alpha$ is
the fine structure constant and $m_e$ and $n_e$ electron mass and
density respectively. We take $n_{e}\approx n_{baryons}$. Under
these conditions $m_{eff}$ is the only information regarding the
mass in the particle's wave equation. Accordingly, we may assume
that $m_{eff}$ behaves mechanically as a true mass and tentatively
apply Caianiello's model to it. Then the weak field approximation
condition $|{\tilde h}_{\mu\nu}|<1$ gives
 \[
 \ r> \sqrt{\frac{2E^2GM}{m_{eff}^3}}\equiv r_\gamma\,.
 \]
Assuming that $ m_{eff}$ makes a contribution when the photon is
well in intergalactic space, but before arrival so that
$d>r>r_\gamma$, we find
\begin{equation}\label{n(e)}
  n_e>\frac{m_e}{4\pi\alpha}\left(\frac{2E^2GM}{d^2}\right)^{2/3}\,.
\end{equation}
Recent measurements of the baryon density by the WAMP
collaboration \cite{WAMP} place upper limits on $n_e$
\cite{raffelt05}. Choosing for illustrative purposes the values
$E\sim 1.4$ eV (for frequencies $\sim 6\,\, 10^{14}$Hz), $M\sim
M_\odot$, and $d\sim 50$ kpc, we find $n_e>9.2\,\,
10^{-10}$cm$^{-3}$, in agreement with the upper limit $n_e<
2.7\,\, 10^{-7}$cm$^{-3}$ from WMAP measurements. This limit gives
$m_{eff}\sim 1.93\,\, 10^{-14}eV$, $r_\gamma\sim 1.5\,\,
10^{-2}d$. Caianiello's model therefore {\it yields the lower
bound} for $n_e$ given by Eq. (\ref{n(e)}) when applied to a
photon of non-vanishing effective mass.

As for the time delay produced by $ m_{eff}$, one finds the
results
\begin{equation}
\delta t_{\gamma}\leq \frac{1}{3
r^{3}_{\gamma}}\left(\frac{2EGM}{m^{2}_{eff}}\right)^{2}\simeq
1.5\,\ 10^{-18}s
\end{equation}
and
\begin{equation}
\delta t_{\gamma k}\simeq t_{o}\frac{m^{2}_{eff}}{2 E^{2}}\sim
1.6\,\ 10^{-16}s\,,
\end{equation}
that are negligible relative to the corresponding neutrino values.

\vspace{0.15in}
\section{Conclusions}
\setcounter{equation}{0}

In conclusion, we have shown that working in the framework of
Caianiello's quantum geometry, the time of flight delay of
neutrinos from SN1987A leads to a lower bound on the neutrino mass
 \[
 m_\nu\gtrsim (10^{-4} - 10^{-3})\mbox{eV}\,.
 \]
This bound is very close to upper bounds coming from cosmological
constraints. Actually, neutrinos with mass $m_\nu \gg k
T_{CMB}\sim 3\,\, 10^{-4}$eV, where $T_{CMB}\sim 3$ K is the
present $CMB$ temperature, contribute to the known mass density of
the Universe $\Omega_m=\rho_m/\rho_c$, where $\rho_c=3H_0^2/8\pi
G_N$ is the critical density and $H_0=100 h$ km/s Mpc is the
Hubble constant ($h\sim 0.7$) relative to non-relativistic matter.
The neutrino density energy $\Omega_\nu$ is given by $\Omega_\nu
h^2=\sum_i m_i/93$eV \cite{dolgov}. Here $\Omega_m h^2\sim 0.15$
and $\Omega_\nu < \Omega_m$. Experimental data about the
cosmological parameters lead to the upper bound on the neutrino
mass sum $\sum_i m_i \lesssim 3$ eV \cite{dolgov}, hence $
m_{\nu}$ should have the upper bound  $m_{\nu}\lesssim 1$ eV.
Future data from MAP/PLANCK CMB and the high precision galaxy
survey (Sloan Digital Sky Survey \cite{sloan}) might relax the
bound to $\sum_i m_i < 0.3$ eV \cite{hu}, or to $\sum_i m_i <
0.12$ eV \cite{han} at $95\%$ C.L.

More stringent limits on the light neutrino mass follow from data
on neutrino oscillations, which fix $\Delta m_{21}^2$ and $\Delta
m_{32}^2$, and from the usual relations that link $m_2$ and $m_3$
to $m_1$
\begin{equation}\label{mass23}
 m_2=\sqrt{m_1^2+\Delta m_{21}^2}\,, \quad m_3=\sqrt{m_1^2+\Delta m_{21}^2+\Delta
 m_{32}^2}\,.
\end{equation}
In the case of normal mass hierarchy $m_1\ll m_2 \ll m_3$ ($\Delta
m^2_{12}=\Delta m^2_{sol}$ and $\Delta m^2_{32}=\Delta
m^2_{atm}$), one finds $m_2\simeq \sqrt{\Delta m^2_{sol}}\sim
7\,\, 10^{-3}$eV and $m_3\simeq \sqrt{\Delta m^2_{atm}}\sim 5\,\,
10^{-2}$, thus $m_1\ll 10^{-3}$eV. Using the oscillation
parameters, the neutrino mass can be expressed in the form
\cite{smirnov}
\begin{equation}\label{massnu}
  m^2_\nu=m_1^2+\left(\sin^2 \theta_{sol}+\cos^2\theta_{sol}|U_{e3}(\theta)|^2\right)\Delta m^2_{sol}+
  |U_{e3}(\theta)|^2\Delta m^2_{atm}\,,
\end{equation}
where, with obvious meaning of the symbols, the matrix element
$U_{e3}(\theta)$ is related to the mixing angle $\theta_{13}$ and
its upper value is $|U_{e3}(\theta)|^2=\sin^2 \theta_{13}\lesssim
5\,\, 10^{-2}$ ($99.73\%$ C.L.). The neutrino mass is then bounded
from above by $m_\nu\lesssim 1.2\,\, 10^{-2}$eV, where the best
fit for the neutrino oscillations parameters give
 \[
 \Delta m^2_{atm}\sim 2.5\,\, 10^{-3}\mbox{eV}^2\,,\quad
 \Delta m^2_{sol}\sim 5\,\, 10^{-5}\mbox{eV}^2\,,\quad
 \tan^2\theta_{sol} \sim 3.4 \,\, 10^{-1}\,.
 \]
One can easily see that the limit can be relaxed to
$m_\nu\lesssim$ a few $\,\, 10^{-3}$eV if $|U_{e3}(\theta)|\sim
3\,\, 10^{-2}$.

For the inverted mass hierarchy $m_2\sim m_3\sim \sqrt{\Delta
m^2_{atm}}$ and $m_1\ll m_{2, 3}$ ($\Delta m^2_{32}\simeq\Delta
m^2_{sol}$ and $\Delta m^2_{21}\simeq\Delta m^2_{atm}$), the
neutrino mass is expressed in the form \cite{smirnov}
\begin{equation}\label{massnuinverted}
  m^2_\nu=m_1^2+(1-|U_{e1}|^2)\left(\sin^2 \theta_{sol}\Delta m^2_{sol}+
  \Delta m^2_{atm}\right)\,,
\end{equation}
where $|U_{e1}| \lesssim 5\,\, 10^{-2}$. The neutrino mass is then
expected to be $m_\nu\sim 5\,\, 10^{-2}$eV.

The $ m_{\nu}$ bounds that we have determined are at least one
order of magnitude smaller than the sensitivity of present
experiments and can not yet be measured in the laboratory. They
are however consistent with what has so far been ascertained. We
point out that the results we have found follow in a natural way
from Caianiello's model, and that they are completely consistent
with the notion that there is a MA linked to the mass of
particles. It should be interesting to investigate the joined
effects of MA and neutrino oscillations within the core of a
supernova, along the line of \cite{dighe}. This goes beyond the
scope of the present paper.

With the caveats discussed in Section III, the model may be
applied to photons with effective mass. One obtains a lower bound
on the electron density in intergalactic space that is compatible
with WMAP measurements.

\vspace{0.2in}

GL and GS aknowledge support for this work provided by MIUR
through PRIN 2004 "Astroparticle Physics", and by research funds
of the University of Salerno. RP's work was performed under the
auspices of the European Union, that has provided financial
support to the "Dottorato di Ricerca in Fisica" of the University
of Salerno through "Fondo Sociale Europeo, Misura III.4". GP
thanks INFN and the Physics Department of the University of
Salerno for continued support and hospitality.

\vspace{0.1in}

\appendix

\section*{}

\vspace{0.1in}

We now calculate the term $a^2$ in (\ref{a^2}) for photons. For
convenience, we perform the calculations using the Schwarzschild
metric. Then, in the plane $\theta=\pi/2$, we obtain
\begin{equation}\label{a^2app}
  a^2=g_{00}{\ddot t}^2+g_{11}{\ddot r}^2+g_{33}{\ddot \phi}^2\,,
\end{equation}
where
\begin{equation}\label{appphi}
  {\dot\phi}\equiv \frac{d\phi}{ds}=\frac{B}{r^2}\,, \quad
  {\dot t}\equiv \frac{dt}{ds}=\frac{A}{1-2MG/r}\,,
\end{equation}
 \[
 {\dot r}\equiv\frac{dr}{ds}={\dot t}\frac{dr}{dt}={\dot t}\left(1-\frac{2GM}{r}\right)
 \left[1-\frac{b^2}{r^2}\left(1-\frac{2GM}{r}\right)\left(1-\frac{2GM}{b}\right)^{-1}\right]^{1/2}\,.
\]
$b$ is the impact parameter and $A$ and $B$ are infinite constants
\cite{wh} when $ds=0$. It is also useful to remember that
\begin{equation}\label{r^2dphi}
  r^2\frac{d\phi}{dt}=r^2\frac{\dot \phi}{\dot
  t}=\frac{B}{A}\left(1-\frac{2GM}{r}\right)
\end{equation}
remains finite as $ds\to 0$ and that $B/A$ is a finite constant
whose value $b/\sqrt{1-2GM/b}$ can be determined by requiring that
$dr/dt=0$ at $r=b$. We show below that the ratio $a^2/{\cal
A}_\gamma^2$, where ${\cal A}_\gamma^2$ is the photon's MA,
remains finite. By using the fact that the components of the
four-velocity in (\ref{appphi}) depend on $r$ only and that
therefore ${\ddot t}={\dot r}d{\dot t}/dr$, ${\ddot r}={\dot
r}d{\dot r}/dr$, and ${\ddot \phi}={\dot r}d{\dot \phi}/dr$, we
can write (\ref{a^2app}) in the form
\begin{equation}\label{a^2app1}
  a^2 \simeq
  \frac{A^4}{r^4}\left[(2GM)^2-4b^2\left(1+\frac{2GM}{b}\right)\right]
\end{equation}
to leading order in $2GM/r$ and $2GM/b$. For purely radial motion,
we find the result of Carmeli \cite{carmeli}
\begin{equation}\label{a^2radialapp}
 a^2\simeq A^4\left(\frac{2GM}{r^2}\right)^2\,,
\end{equation}
from which it also follows that ${\cal A}^{2}_\gamma=A^4/(2GM)^2$
and that the ratio
 \[
  \frac{a^2}{{\cal A}^2_\gamma}\leq \left(\frac{2GM}{r}\right)^4
 \]
remains finite. We finally obtain from (\ref{a^2})
\begin{equation}\label{ultima}
 {\tilde h}_{\mu\nu}k^\mu k^\nu=\frac{a^2}{{\cal
 A}^2_\gamma}\eta_{\mu\nu}k^{\mu}k^{\nu}\leq k^2\left(\frac{2GM}{r}\right)^4=0
 \end{equation}
for on-shell photons.

\newpage

\begin{table}\label{table}
\caption{Observation times and neutrino energies in the Kamioka
experiment [1,4].}
\begin{tabular}{ccc}
  Event No. &  $t_{obs}$(s) & Energy(MeV)  \\ \tableline
  1 & 0 & $21.3\pm 2.9$ \\
  2 & 0.107 & $14.8\pm 3.2$ \\
  3 & 0.303 & $8.9\pm 2.0$ \\
  4 & 0.324 & $10.6\pm 2.7$ \\
  5 & 0.507 & $14.4\pm 2.9$ \\
  6 & 0.686 & $7.6 \pm 1.7$ \\
  7 & 1.541 & $36.9\pm 8.0$ \\
  8 & 1.728 & $22.4\pm 4.2$ \\
  9 & 1.915 & $21.2\pm 3.2$ \\
  10 & 9.219 & $10.0\pm 2.7$ \\
  11 & 10.433 & $14.4\pm 2.6$ \\
  12 & 12.439 & $10.3\pm 1.9$ \\
\end{tabular}
\end{table}

\newpage

\begin{figure}
\centering \leavevmode \epsfxsize=9cm \epsfysize=7cm
\epsffile{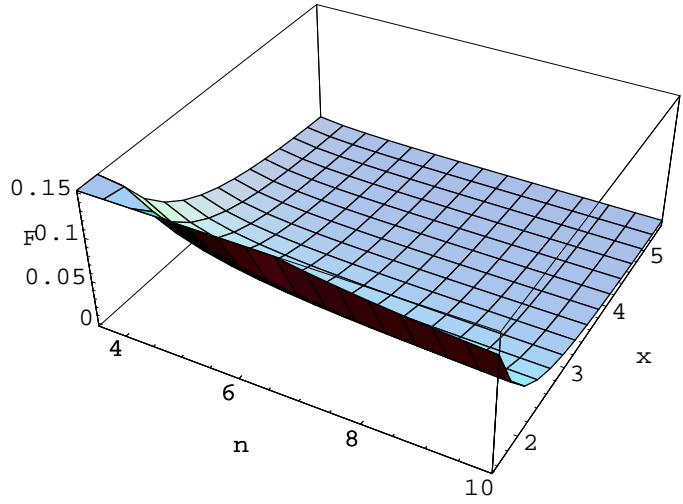} \caption{\small{Behaviour of $F(x,n)$ for
$x\;\epsilon[1.5,3.5]$ and $n\;\epsilon[3.5,10]$.}} \label{F(x,n)}
\end{figure}

\newpage

\newpage

\end{document}